# Business Process Measures

Valdis Vitolins

University of Latvia, IMCS, 29 Raina blvd, LV-1459, Riga, Latvia
valdis_vitolins@exigengroup.lv

**Abstract**. The paper proposes a new methodology for defining business process measures and their computation. The approach is based on metamodeling according to MOF. Especially, a metamodel providing precise definitions of typical process measures for UML activity diagram-like notation is proposed, including precise definitions how measures should be aggregated for composite process elements.

**Keywords.** Business process, model, metamodel, measure.

## 1. Introduction

Globalization and increasing competition forces companies to improve their business, but improvements can't be evaluated without measurements. Thus, measures of business processes in companies provide crucial knowledge of TCO and ROI for executives. Steady growth of IT usage in business makes these measurements more feasible [1].

Many quality [2,3] and business process management [4,5,6] methodologies now use numeric methods to figure out weaknesses and strengths of a business. Methodologies are supported by several tools [7,8,9,10], however they provide the "best of breed" methodology only for one narrow area and they can't support several methodologies simultaneously.

In this paper the research on business process modeling problems [11] is continued. Measures and rules, they relation to business concepts are shown in a formal and unambiguous way, using Unified Modeling Language (UML) [12]. Theories and methodologies are analyzed using metamodeling approach according to Meta Object Facility (MOF) [13].

Gradually, through given examples for each abstraction (meta) layer, several aspects of business process measures are precisely defined. It is shown that on the one hand, each higher abstraction layer describes concepts that are more common, but on the other hand, each higher layer determines rules and possibilities for lower layers.

The model (M1 layer) is represented by a business process example (UML activity diagram extended by measures) and a class diagram defining the "measure view" of the same example. The most significant and more detailed is metamodel (M2 layer), because it determines common possibilities and features for business process models in a modeling notation. Extended metamodel shows how standard measures and measure aggregation facilities for composite objects are defined. The classification of reasonable process measures and their possible assignment to process elements is provided. The metametamodel (M3 layer) briefly sketches a universal framework for measure definition, from which the specific "measuring metamodel" (M2) could be obtained as an instance.

The proposed metamodel can be used as a unified framework for the development of comprehensive business modeling and measurement tools.

## 2. Business Process Model (M1)

To demonstrate measuring of a business process on a practical example, a specific modeling language will be used. This language is a slightly modified UML Activity diagram (AD) [12] with **extensions for object measures**. Mainly the graphical notation of activity diagram is made more expressive and the terminology is changed, but the semantics is a standard one, except the resource management, which is made more precise. The language corresponds also to the previously developed business process metamodel [11]. References to diagram elements are shown in *italic* in the text. The example (Fig. 1) shows one business process for a shop, which delivers pizzas to customer homes. *Sell Pizzas* is a business process (an activity in AD notation), which consists of several tasks (actions in AD).

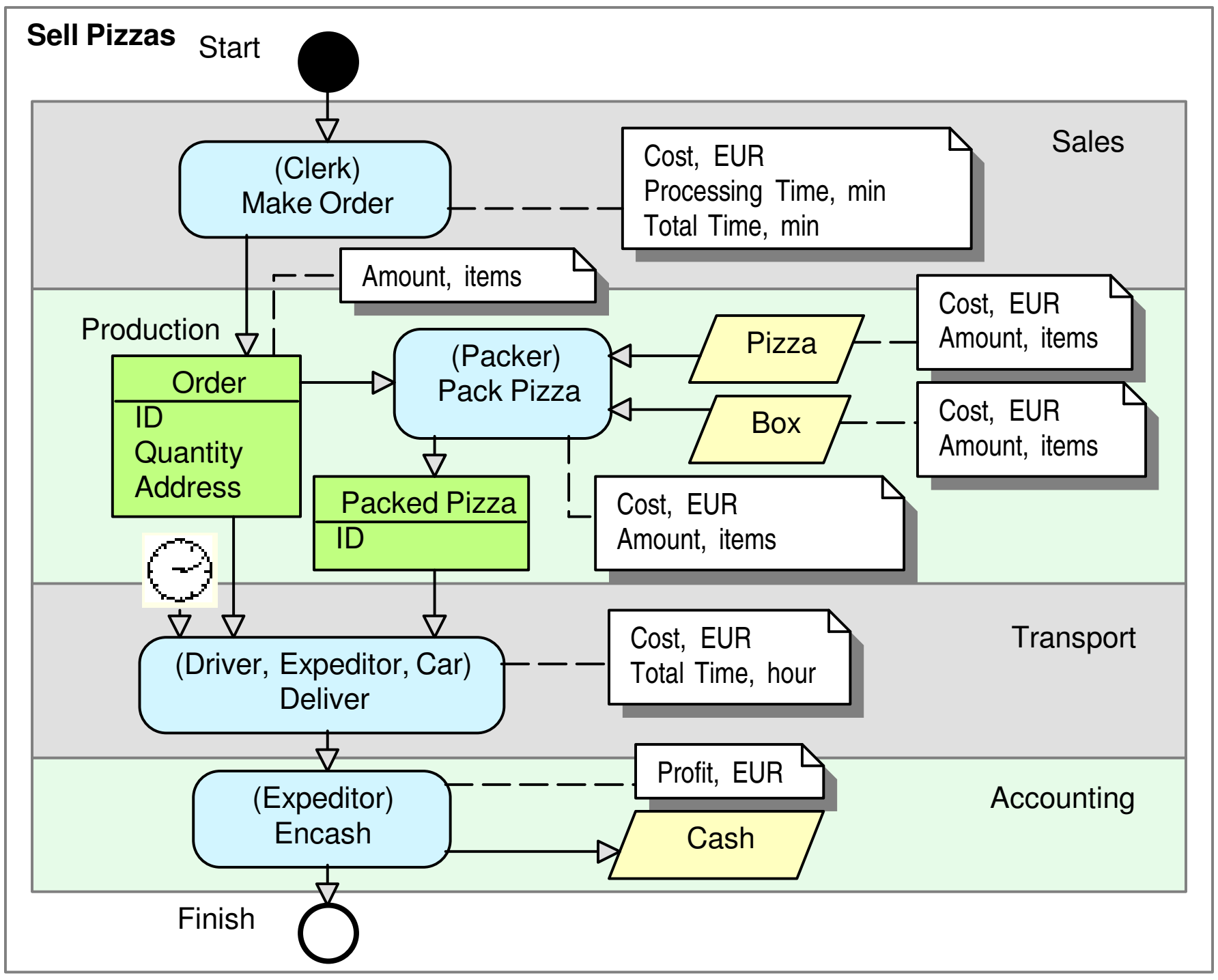

**Fig. 1** Sell Pizzas business Process (M1)

Tasks are represented by rounded rectangles. Some tasks have **performers** in parentheses. Performer is a special reference that is used to specify that the given employee or resource is necessary to perform the task. Control flows are shown by simple arrowed lines, and rectangles represent objects in object flows – either flows of messages (e.g. *Order*) or physical objects (e.g. *Packed Pizza*) with their attributes. Parallelograms correspond to AD datastores, where materials are located. An object flow entering a datastore means that putting into the datastore, but leaving flow - taking. The fact that *Deliver* is a task, which is started at some regular time moments is shown by a symbolic clock (Time event in AD). By default, all incoming flows are joined at a task with AND condition. Organizational units that are responsible for each task (e.g. *Sales*) are shown as swim lanes. If a specific performer for a task isn't set, anyone from the corresponding organizational unit can perform the given task. If performer is set, only the given performer can perform the given task. Each process (or subprocess) starts with *Start* node and finishes with *Finish* node.

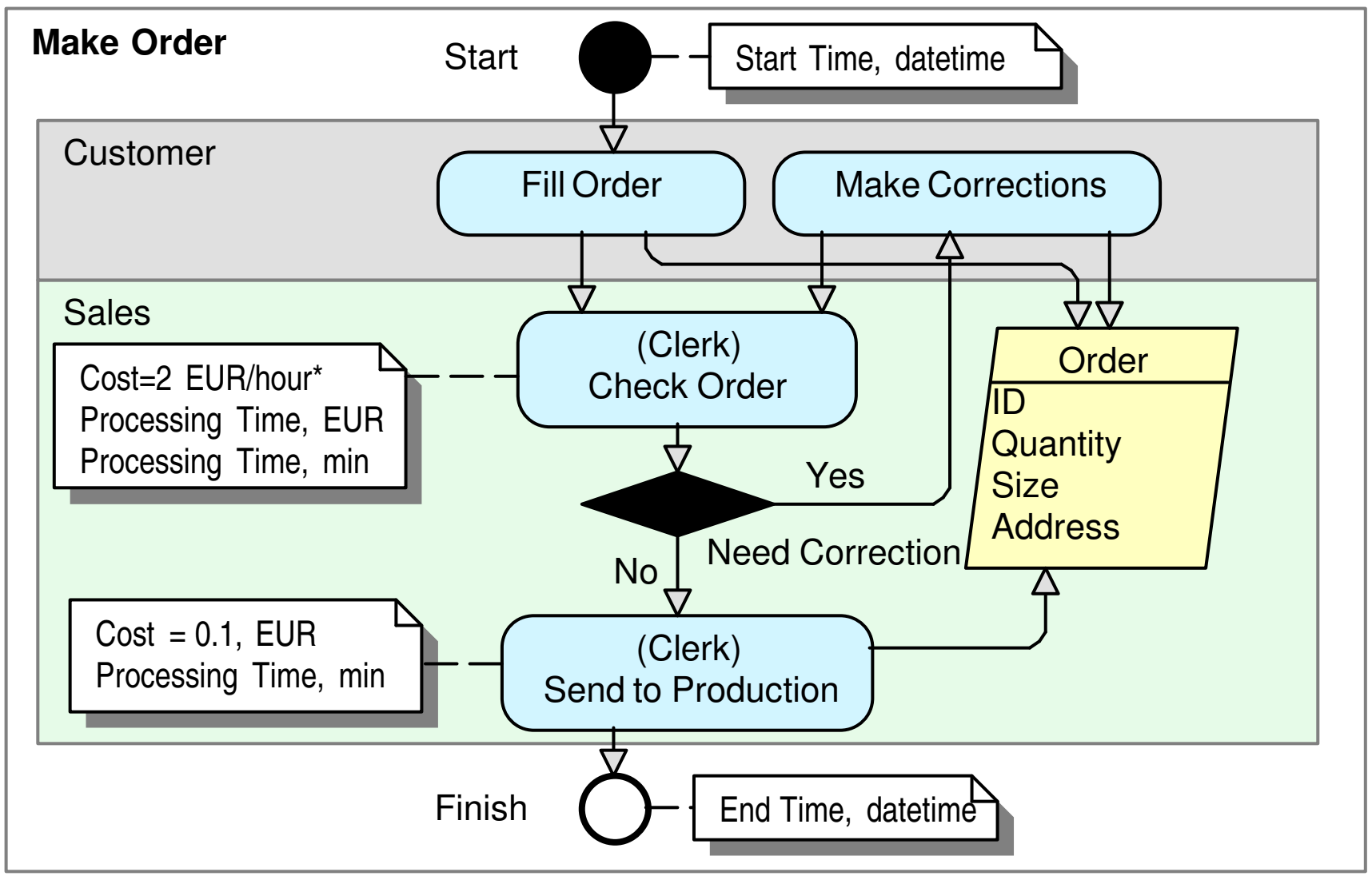

**Fig. 2** Make Order business process (M1)

*Make Order* in Fig. 2 actually is another business process which is decomposes the task *Make Order*. This subprocess contains a decision *Need Correction* (diamond). The complete *Sell Pizzas* business process is shown only for a "full picture" and to demonstrate that several measures for some business process objects can be assigned. **For further explanation, only the *Make Order* subprocess and *Make Order* task will be used**.

Measures are shown in notes (e.g. *Cost*, *EUR*), which are linked to the measured element (object) by a dashed line. Each measure in a note is specified by its name, an optional declaration and the unit according to the following syntax: Name[=declaration],Unit (e.g. *Cost=2 EUR/hour*Processing Time, EUR*). The measure declaration can be empty, a constant or expression. If a measure declaration references a name of other measure, by default the measure for the same object is assumed. From analysis of several process management tools [7, 9, 10], it was assumed that each system provides information about system time and performers. Therefore, measures, which rely directly on system time and performers don't need declarations. All other measure values should be declared implicitly in metamodel, or explicitly as constants or formulas in model.

A measure linked to a process element means that the given measure must be evaluated during the process execution. See more on measure values in sections 3 and 9. According to MOF this business process model is an abstraction of all execution instances of the real business process therefore it conforms to M1 layer.

For clear understanding of all metalayers, an example of instance layer M0 also will be shown.

## 3. Measure Value Instances (M0)

During the system run/simulation of the business process, shown in Fig. 1 and Fig. 2, each of the defined objects in the model gets its runtime instance. Each runtime instance of an object has all its attribute values set, including the values of measures linked to the object. In Tab. 1, instances are shown as a "denormalized" table (view), where **rows correspond to the registered measure instances**.

Instances from the *Make Order* sub-process are shown in this table for one process execution. Columns *Object Type* and *Object* represent the owning business object instances. Cells in the *Measure Declaration*, *Unit* column represent the identification of each measure. All the above-mentioned cells are actually resolved from business process model (they are not instance dependent). Cells in *Value* and *Time* columns represent values for each measure instance. Cells in these columns are filled by the process management system according to the actual process execution. *No*, *Source No* columns are for information only and describe how derived values are calculated.

The table illustrates that some values are got explicitly from the management system (e.g. value for *Processing Time*), but some values are calculated implicitly through given rate and explicit value (e.g. *Cost* from *2 EUR/hour* and *Processing Time*), or from values of several sub-measures (e.g. *Cost* for *Make Order*, from two separate costs).

| No | Object Type | Object | Measure Declaration, Unit | Value | Time | Source No |
|---|---|---|---|---|---|---|
| 1 | Business Process | Make Order | **Start, datetime** | **9:05:34** | **9:05:34** | - |
| 2 | Task | Check Order | **Processing Time, min** | **0:03:00** | **9:10:03** | - |
| 3 | Task | Check Order | **Cost=2 EUR/hour*Processing Time, EUR** | **0,10** | **9:10:04** | 2 |
| 4 | Task | Send to Production | **Processing Time, min** | **0:02:00** | **9:15:40** | - |
| 5 | Task | Send to Production | **Cost =0.1, EUR** | **0,10** | **9:15:45** | - |
| 6 | Business Process | Make Order | **Finish, datetime** | **9:15:47** | **9:15:47** | - |
| 7 | Business Process | Make Order | **Processing Time, min** | **0:05:00** | **9:15:47** | 2,4 |
| 8 | Business Process | Make Order | **Total Time, min** | **0:10:13** | **9:15:47** | 1,6 |
| 9 | Business Process | Make Order | **Cost, EUR** | **0,20** | **9:15:47** | 3,5 |
| 10 | Business Process | Make Order | **Processing Time, min** | **0:05:00** | **9:15:48** | 2,4 |

**Tab. 1** Sample Values for Make Order Sub-process

General rules according to which this table could be obtained from the "raw material" – a complete process execution log are described in section 9.

## 4. Measure Aggregation Sample Model (M1)

Though the activity diagrams in Fig. 1 and Fig. 2 represent the M1 layer, they are only “graphic interfaces” for the control aspects of the business process model, and in such representation not all measure-related items are viewable. Therefore, the measure aspect of the same model is shown in full details, using a class diagram (Fig. 3). This class diagram is another view for the same process, where “system” objects that actually exist and are necessary for process measuring are made explicit. At the same time, the control and execution aspects of the model are not visible in this view.

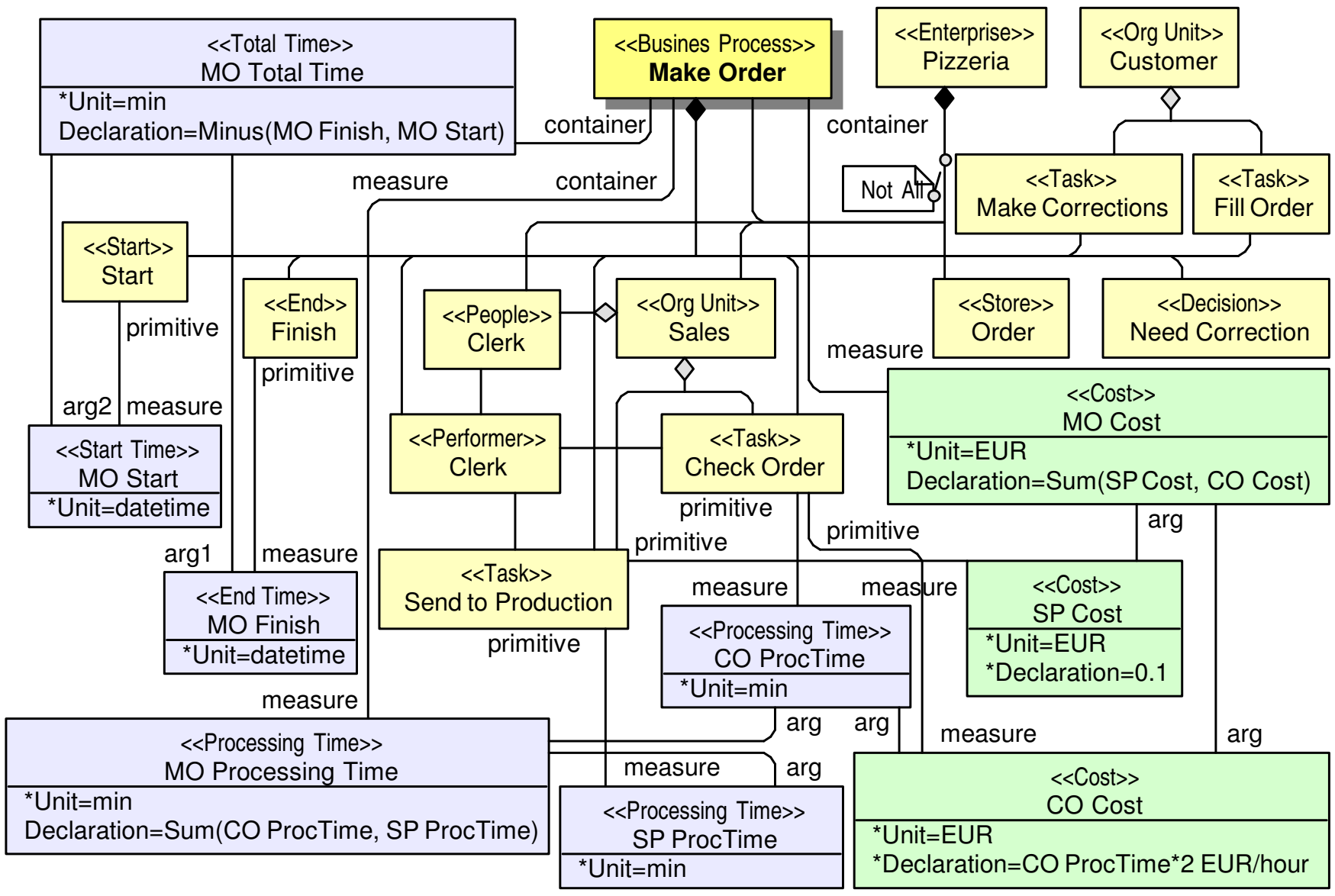

**Fig. 3** Measure Sample Model (M1)

The diagram is built in accordance with the process measuring metamodel (M2 level) in Fig. 5. E.g., object relations to their measures are shown as relations with the specified role names *primitive/container-measure*. If a measure declaration references another measure, it is shown with role name *arg*.

In this diagram on the one hand each class is one specific **instance** of a more abstract class in M2 layer (Fig. 5), but on the other hand it is a **class**, because it represents all possible instances from the M0 metalayer (Tab. 1). So this means, that they all are classes with more specific properties than classes in M2 metalayer. Let name such classes “instance classes” (it is an extension of UML class diagram notation). This is shown by using stereotypes in such a way, that a class name in a higher metalevel (e.g. M2) becomes to the stereotype for the corresponding instance classes in the lower metalevel (e.g. M1).

According to the metamodeling traditions in MOF (including the metamodel for UML class), components that compose a measure at M2 metalayer, are shown as new specific tagged compartments of a concrete measure class at M1 layer (e.g.

*unit=EUR*, *declaration=Minus(MO_Finish,_MO_Start)*). Compartments, what are explicitly defined in business process model (Fig. 1, Fig. 2), are marked with asterisk. Unmarked compartments are derived implicitly from measure declaration metamodel.

Unfortunately, a universal application of this principle doesn't work well always. E.g., if this rule would be used for *Business Process* class, it should be shown as a single class with compartments at M1 layer. In such way, diagram would become too unreadable, therefore, some decompositions at M2 layer are treated as decompositions at M1 layer. E.g., *Business Process* and *Enterprise* are shown as usual decomposition with separate classes and corresponding stereotypes.

## 5. Business Process Metamodel (M2)

In Fig. 1 a model of one particular business process (M1) was shown, but for the development of a universal process measuring method, an adequate business process metamodel (Fig. 4) is necessary (M2 layer), as the base for that notation.

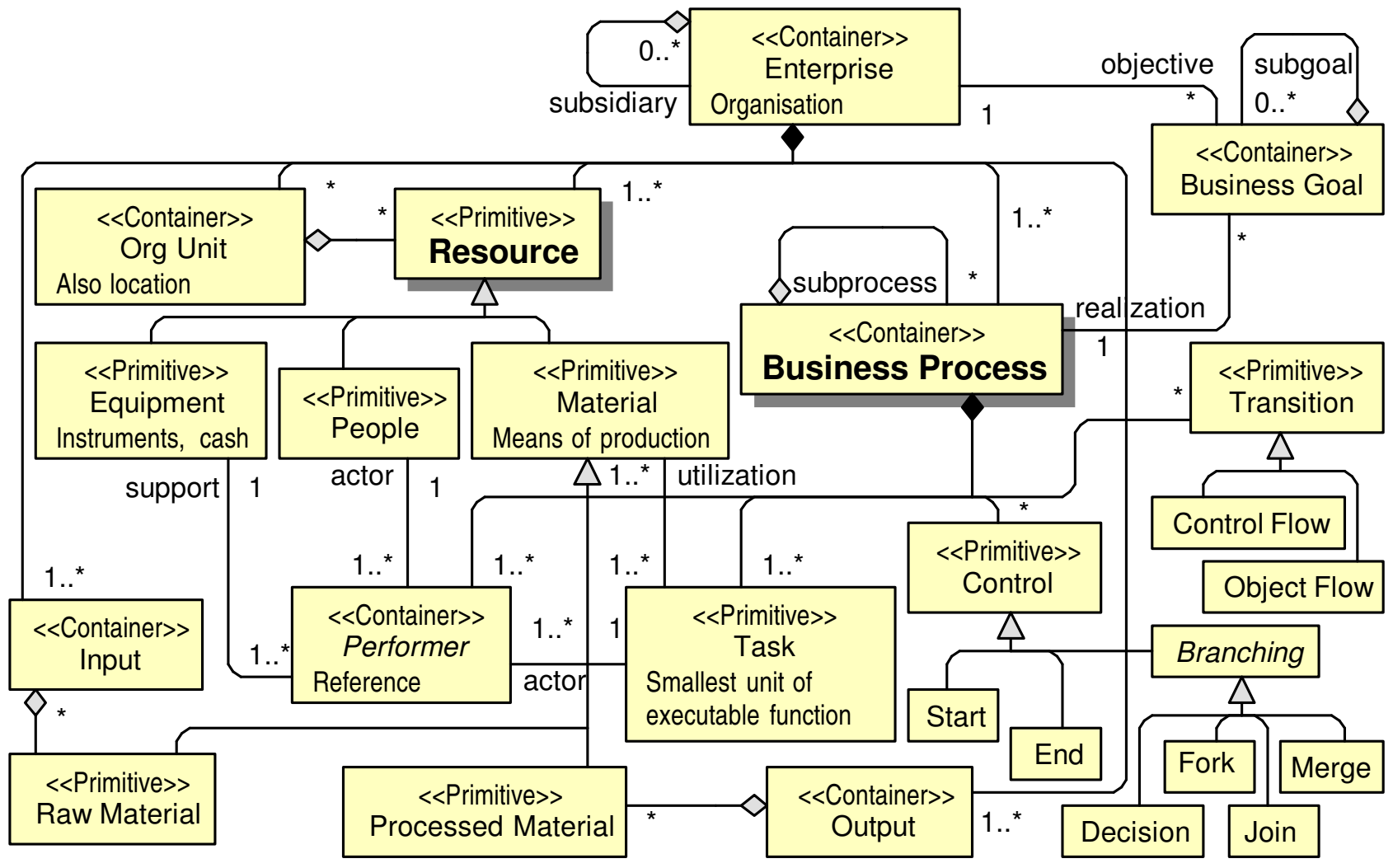

**Fig. 4** Business Process Metamodel (M2)

A common view on business processes has been already studied in [11]. In the current research, the main interest is directed towards the dynamic behavior of an enterprise, therefore, other concepts are included, only if they play a role for business processes. It actually is a cut (and renamed) version of the UML activity diagram metamodel, but with the resource management aspect added.

In this metalayer all classes are "instance classes" again, because they are specific instances for the more abstract metametalayer (M3, Fig. 7).

A business process is developed as a *realization* of some *Business Goals* in *Enterprise*. Each of the Enterprise has *Resources*, *Organizational Unit* and *Business Processes*. *Material* is one of *Resources* that belongs to *Organizational Unit*. *Raw*

*Material* flows from *Input* of *Enterprise* to *Output* and becomes to *Processed Material* through *utilization* in a *Task*. *Task* is an atomic part of *Business Process* and doesn't contain anything else (i.e. elementary action in UML AD). A *Task* has a special reference - *Performer*, which is used for indirect pointing to a *Resource* (such as *Equipment* and *People*), that performs or is necessary for the task execution. The amount of *Material,* that is utilized in a task, is decremented after the task execution. The count of free *People* and *Equipment* is used only (is decremented) during task execution time, and becomes free (is incremented) after.

*Controls* are used for controlling the task execution sequence (branching - parallel processing and decisions). *Transition* (*Control Flow*) shows task execution sequence, but *Object Flow* additionally shows what object (message or physical material) is sent from task to task (as in UML AD).

## 6. Measure Definition Metamodel (M2)

In addition to the described business process metamodel (section 5), a harmonized measure declaration metamodel (M2 layer) is derived. This metamodel (Fig. 5) provides standard declarations for the most typical measures. This model again uses the “instance class” notation with respect to the metameta layer (Fig. 7). The main value of this metamodel is in the fact that it shows how process element structuring translates into appropriate measure value aggregation for the most typical measures.

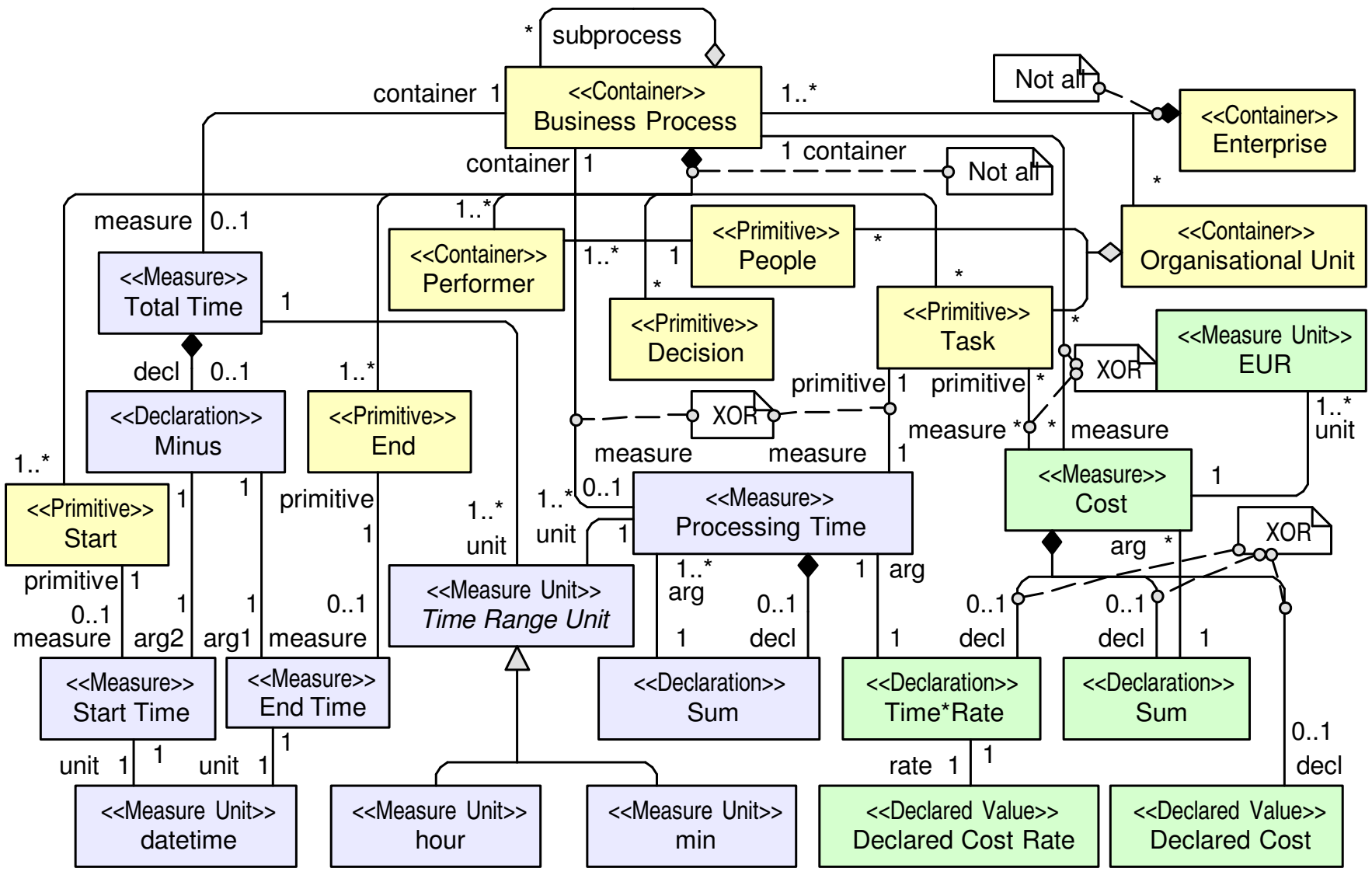

**Fig. 5** Measure Declaration Metamodel (M2)

In the metamodel it is defined, how the selected measures are related to business process objects and to each other. If a measure has a declaration expression (an implicit one or to be explicitly defined in a model), this expression is shown as a function-arguments complex, the function having the <<*Declaration*>> stereotype. In Fig. 5, it is shown, which business process elements can have *Start Time* and *End Time* measures. Some process elements e.g., *Business Process* or *Task* can have several measures, such as *Processing Time*, *Total Time* or *Cost*. A declaration for a measure depends on the process element to which it is attached and how it is declared (when explicit declaration in a model is possible). E.g., *Cost* can be declared either as an exact value (*Declared Cost*), as a multiplication of *Processing Time* by *Declared Cost Rate*, or as a sum of several sub-costs (the only possibility if the *Cost* measure is attached to a <<*Container*>> object - e.g. *Business Process*).

*Processing Time* can be without declaration (if the value is obtained directly – it is attached to a *Task*), or can be a sum of several *Processing Times* (if the measure is attached to a *Business Process* object). A measure can have several *Units*. In this example, several units (e.g. *hour*, *min*) are shown only for the *Processing Time* measure.

This metamodel could serve as basis for completely precise measure declaration scheme, where all implicit declarations and possible explicit declarations for each feasible measure-object pair can be specified. Namely, the pair object-measure makes the declaration features unique, measure classes in Fig. 5 are "reused" to reduce the diagram size. All this could be specified formally as standard OCL [14] constraints at measure classes.

## 7. Measure Classification and Relation to Business Objects (M2)

Through expanding of the partial measure declaration metamodel (Fig. 5), a complete metamodel is obtained. Possible assignments of measures to business process elements (from Fig. 4) are shown in Fig. 6.

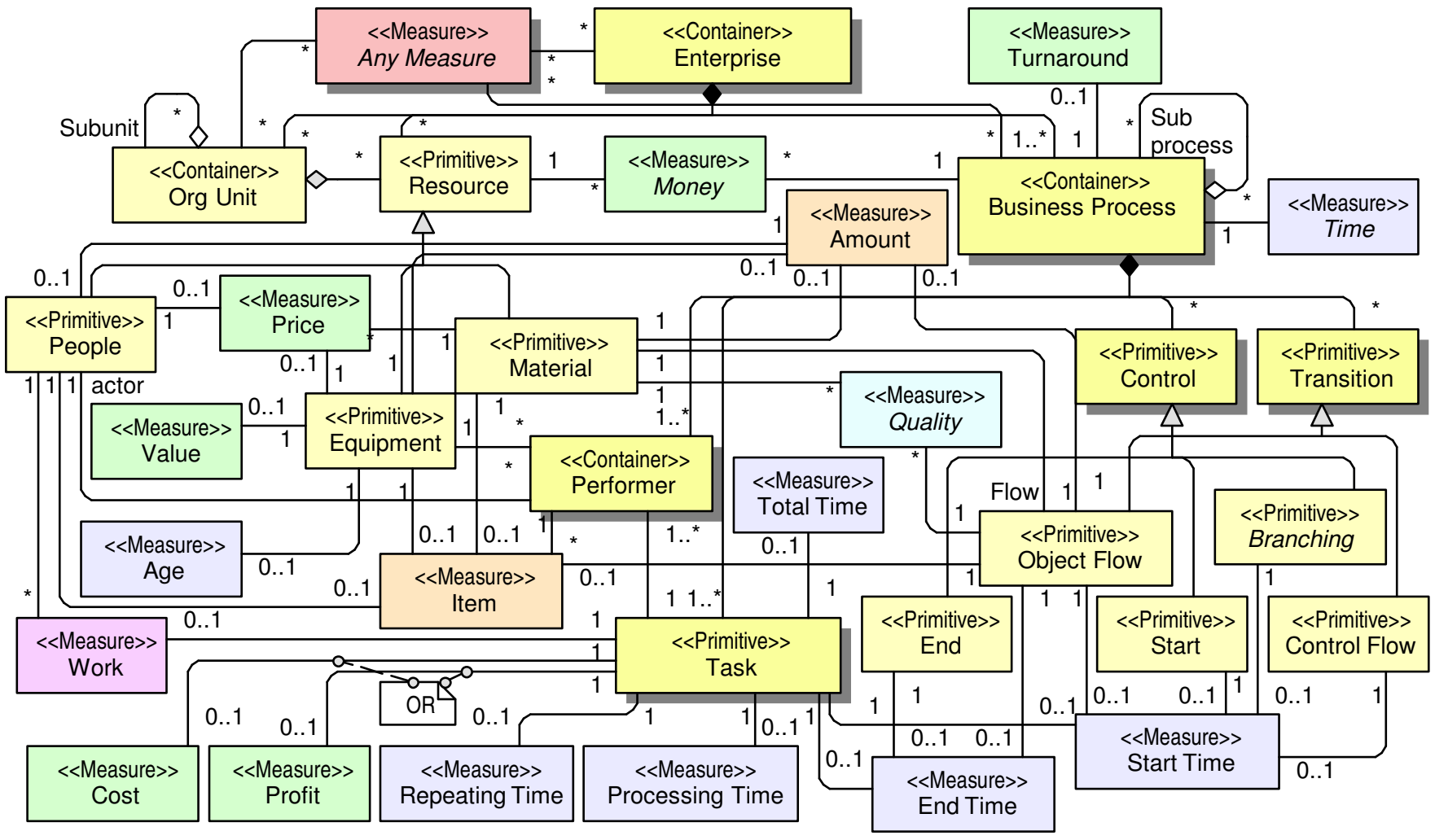

**Fig. 6** Measure Relations to Business Objects Metamodel (M2)

Measures used in typical process modeling tools [7,9,10] actually are subsets of the assignment provided here. Therefore the metamodel in Fig. 6 could serve as a certain standard. However, since another useful measures can be always be invented, a general framework for assigning any measure in a similar way to any process element (business object) will be described in section 8.

Through analysis of several management approaches [2,3,4,5,6], the following measure groups were introduced: *Time*, *Money*, *Resource*, *Work* and *Quality* (marked as blue, green, orange, violet and azure colors). Unnamed associations are used in Fig. 6 to assign a measure (a class with the stereotype <<*Measure*>>) to a business object (<<*Primitive*>> or <<*Container*>>).

The provided metamodel has a significant value, since attaching any measure to any process element would be a semantic nonsense. Measure declarations, constraints and aggregation functions are out of scope of the current paper, therefore it will not be described in details here.

## 8. Business Measure Metametamodel (M3)

To measure and analyze business in a comprehensive way, a generic and common methodology for all possible business management areas is necessary. Therefore one more abstraction layer or metametamodel (M3) should be introduced. I.e., building a tool on the basis of a more abstract layer provides the possibility to add new measures for objects or change existing.

According to MOF traditions, the metametamodel (the MOF M3 layer) is kept simple and fixed, and all the complexity of a specific domain should be represented by its metamodel. This would imply an intensive use of OCL constraints for a metamodel of a complicated domain, in order to specify the intended semantics. Here another solution has been tried (a legal one with respect to MOF standards), the metametamodel is extended by specialized classes, in order to specify semantic constraints for a set of metamodels in a readable way.

In the proposed metametamodel (Fig. 7) the new metaclasses *Business Object* and *Measure* are defined as specializations of *Class* from UML *InfrastructureLibrary::Constructs*. The *child* association is meant to be the same one from UML metametamodel, which associates a class as a part of another (actually the real metametamodel is more complicated there). In this way we can retain the UML metametamodel for the “modeling part” of metamodels, and have a framework for modifying types of measures. The main aspect we want to know from this specific metametamodel is, which metaclasses (business objects) represent the primitive model elements and which the container (composite) ones. The metamodels (Fig. 4,5,6) can be obtained as “instantiations” of the proposed metametamodel, using the respective <<*Primitive*>> or <<*Container*>> stereotypes for business objects.

According to the metametamodel (Fig. 7), a *Measure* is a concept, whose main part is its *Declaration*. The measure *Declaration* **can** contain a *Declared Value* (a constant), or several *Math Functions*, which use *Declared Values* or other *Measures* as argument (*arg* role). If a *Business Object* is a *Container*, the corresponding

*Measure* **must** contain *Aggregation Function*, which uses other *Measures* that belong to the contained *Business Objects* as a source.

A measure *Declaration* can reference another measure in two ways – either explicitly a named measure (e.g., *Processing Time* in Fig. 2), or implicitly through *Aggregation Function*, if the measure is attached to a *Container* object (e.g., *Cost* for *Make Order* in Fig. 1). A *Measure* references also one of several possible *Measure Units*.

Each measure is associated to some *Business Object* (a relevant element of the business process metamodel). *Business objects* can be either *Primitive* (e.g. Task), or *Container* (e.g. *Business Process*). If a business object is a *Container*, and its measure has no *Declared Value*, then values of this measure **must** be calculated using the specified values from *child* (contained) objects.

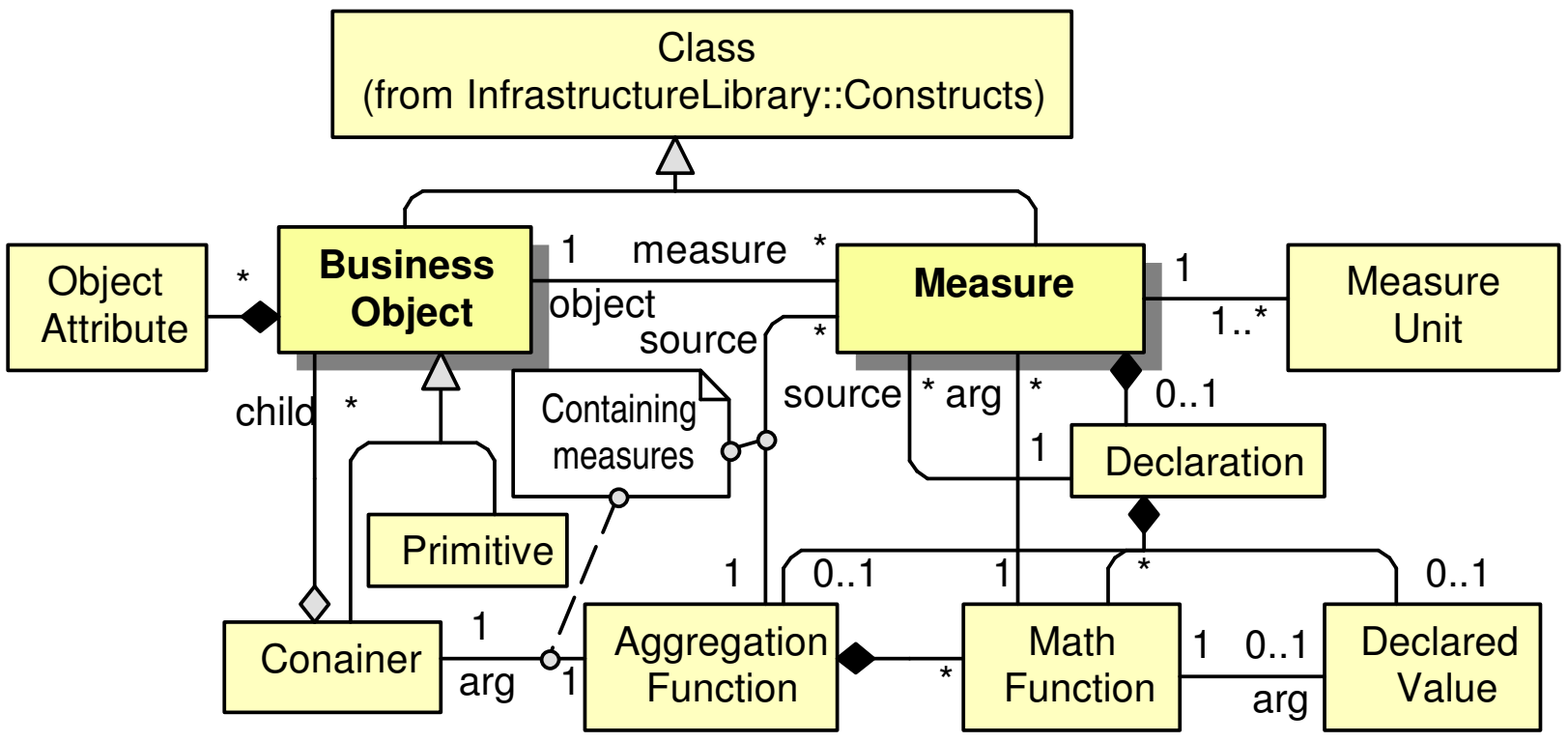

**Fig. 7** Measure Metametamodel (M3)

## 9. Business Process Execution and Measure Values

**Measures** itself are only declarations or definitions of what and how some business objects will be measured. As it was described in section 4, actual values are obtained only at system runtime. For declaration execution, an exact semantics dependant only on the measure declarations, but not on the way the process is run, should be defined.

Existing process management and simulation tools each have a different definition of process semantics. On the basis of a common framework for business process modeling [11] a common framework for process execution semantics could also be defined, but details are well beyond the limits for this paper.

Assume that the some execution engine executes processes, shown in Fig. 1, 2, and provides execution log shown in Tab. 2. In the following table, in contrast to Tab. 1, **row represents a dynamic instance of an each model element** (i.e. including instances of "system" elements). Cells in columns *Object Type* and *Object* represent the object (diagram element). Cells in *Start Time*, *End Time*, *Processing Time* columns show the actual values of instance attributes. *No* column is a unique instance (row) identifier in the table.

The most important for value aggregation is the *Process ID* column (new Token ID or a new process copy identifier). The new unique ID is generated for each **new** process execution, when business process starts in its start point in the top level business process. In the proposed notation the start point is declared explicitly, but it could also be declared implicitly. This process ID is used for all process elements till the process instance end.

| No | Object Type | Object | Process ID | Start Time | End Time | Processing Time | Performer |
|---|---|---|---|---|---|---|---|
| 10023 | Business Process | Sell Pizzas | 00101 | 9:05:34 | 10:15:35 | | |
| 10024 | Business Process | **Make Order** | 00101 | **9:05:34** | 9:15:47 | | |
| 10025 | Task | Fill Order | 00101 | 9:05:34 | 9:06:03 | 0:01:31 | |
| 10026 | Business Process | Sell Pizzas | 00102 | 9:06:12 | 10:20:12 | | |
| 10027 | Business Process | Make Order | 00102 | 9:06:13 | 9:08:02 | | |
| 10028 | Task | Fill Order | 00102 | 9:06:13 | 9:07:01 | 0:01:12 | |
| 10029 | Task | Check Order | 00102 | 9:07:01 | 9:07:48 | 0:00:47 | Clerk2 |
| 10030 | Decision | Need Correction | 00102 | 9:07:48 | | | |
| 10031 | Task | Make Corrections | 00102 | 9:08:57 | 9:16:05 | | |
| 10032 | Task | **Check Order** | 00101 | **9:10:03** | 9:13:03 | **0:03:00** | Clerk1 |
| 10033 | Decision | Need Correction | 00101 | 9:15:39 | | | |
| 10034 | Task | **Send to Production** | 00101 | **9:15:40** | 9:17:40 | **0:02:00** | Clerk1 |
| 10035 | Task | Check Order | 00102 | 9:16:10 | 9:17:48 | 0:01:38 | Clerk2 |
| 10036 | Decision | Need Correction | 00102 | 9:17:48 | | | |
| 10037 | Task | Make Corrections | 00102 | 9:17:59 | 9:20:00 | | |
| 10038 | Task | Check Order | 00102 | 9:16:10 | 9:17:48 | 0:01:38 | Clerk2 |
| 10039 | Decision | Need Correction | 00102 | 9:16:12 | | | |
| 10040 | Task | Send to Production | 00102 | 9:20:01 | 9:21:02 | 0:01:01 | Clerk2 |

**Tab. 2** System Runtime Log Example

This execution log is used for extracting measure values. Rows in Tab. 2 that are used for the generation of the measure value example in Tab. 1 are marked yellow, and the attribute values that are really used for the measure value calculation are shown in bold rectangles.

For container objects such as business processes the measure aggregation is used according to the implicit declarations specified in Fig. 3 and definition in Fig 2. The aggregation is always performed for the same process execution instance, using *Process ID*.

The described principles are sufficient for a formal definition of a universal measure value extraction procedure. Such a procedure would provide the precise measuring semantics. However, there may be more complicated situations within the described framework. E.g., if both the main and subprocess diagrams contain loops, the provided identification at the process execution instance level is insufficient. The finding of best universal value extraction procedures is a theme for future research.

# 10. Conclusions

Business process management systems or even simulation experiments produce large amount of plain data, which show no clear picture about the real business process. Aggregation and analysis of these data requires development of new methods and calculations, because existing tools support only a small part of interests.

In the current paper, a new look on business process measurement problem is proposed. The problem is analyzed in an unambiguous and formal way using UML. Several process measuring methodologies are merged with the metamodeling approach according to MOF, and a comprehensive business process measurement metamodel has been developed.

The proposed approach allows defining values in a natural way, and measurement of data, which are of interest to business, without deep investigation into specific technical solutions. This provides new possibilities for business process measurement, decreasing the gap between technical solutions and asset management methodologies.

As a further research, development of a more detailed metamodel and standardization of system runtime is planned. The research results will provide a framework for metamodel-based business modeling/simulation/management tools, and will extend them with comprehensive business process measurement possibilities.